\documentclass[preprint,eqsecnum,showpacs,nofootinbib,aps,amsmath,amssymb,tightenlines]{revtex4}

\usepackage{amsmath,amssymb,amsbsy,latexsym,graphics}

\def\ut#1{\rlap{\lower1ex\hbox{$\sim$}}{#1}}

\def\R{{\mathbb R}}
\def\S{{\mathbb S}}

\def\M{{\mathbb M}}
\def\J{{\mathbb J}}

\def\D{{D}}
\def\man{{\cal M}}

\def\Lie{{\cal L}}

\def\IH{{\Delta}}
\def\={\,\widehat{=}\,}
\def\l{\ell}
\def\t{\tilde}
\def\tS{\t{S}}
\def\tq{\t{q}}
\def\tw{\t{\omega}}

\def\tE{\t{E}}
\def\tB{\t{B}}
\def\tR{\t{\cal R}}
\def\h{\hat}

\def\k{\kappa}
\def\d{{\rm d}}
\def\f{\frac}

\def\ba{\begin{eqnarray}}
\def\ea{\end{eqnarray}}
\def\be{\begin{equation}}
\def\ee{\end{equation}}

\def\Re{\mathrm{Re}}
\def\Im{\mathrm{Im}}

\newcommand{\dual}{{}^\star}

\preprint{\vbox{\baselineskip=12pt
 \rightline{gr-qc/0401xxx}
 \rightline{CGPG 2004-01/3}
\rightline{NSF-KITP-04-12} }}

\begin{document}


\title{Multipole Moments of Isolated Horizons}

\author{Abhay\ Ashtekar${}^{1,3,4}$}
\email{ashtekar@gravity.psu.edu}
\author{Jonathan\ Engle${}^1$}
\email{engle@gravity.psu.edu}
\author{Tomasz Pawlowski${}^{1,2,4}$}
\email{Tomasz.Pawlowski@fuw.edu.pl}
\author{Chris Van Den Broeck${}^{1}$}
\email{vdbroeck@gravity.psu.edu}

\affiliation{${}^1$Center for Gravitational Physics and Geometry,\\
    Physics Department, Penn State, University Park, PA 16802,
    USA\\
    ${}^2$Institute for Theoretical Physics\\
    Warsaw University, ul. Hoza 69, 00-681, Warsaw, Poland\\
     ${}^3$Kavli Institute of Theoretical Physics\\
    University of California, Santa Barbara, CA 93106-4030, USA\\
    ${}^4$Erwin Schr\"odinger Institute, Boltzmanngasse 9, 1090 Vienna,
    Austria\\}


\begin{abstract}

To every axi-symmetric isolated horizon we associate two sets of
numbers, $M_n$ and $J_n$ with $n = 0, 1, 2, \ldots$, representing
its mass and angular momentum multipoles. They provide a
diffeomorphism invariant characterization of the horizon geometry.
Physically, they can be thought of as the `source multipoles' of
black holes in equilibrium. These structures have a variety of
potential applications ranging from equations of motion of black
holes and numerical relativity to quantum gravity.
\end{abstract}


\maketitle
\section{Introduction}
\label{s1}

Multipole moments play an important role in Newtonian gravity and
Maxwellian electrodynamics. Conceptually, there are two distinct
notions of multipole moments ---source multipoles which encode the
distribution of mass (or charge-current), and field multipoles
which arise as coefficients in the asymptotic expansions of
fields. In Newtonian gravity, the first set is of direct interest
to equations of motion of extended bodies while the second
determines the gravitational potential outside sources. However,
via field equations one can easily relate the two sets of
multipoles.  In the Maxwell theory, the rate of change of the
source dipole moment is directly related to the energy flux
measured at infinity. Because of such useful properties, there has
been considerable interest in extending these notions to general
relativity.

Results of the Maxwell theory were extended to the weak field
regime of general relativity ---i.e., linearized gravity--- quite
some time ago. Already in 1916 Einstein obtained the celebrated
formula relating the rate of change of the quadrupole moment of
the source to the energy flux at infinity \cite{ae}. In the
fifties, Sachs and Bergmann extended the relation between the
source multipoles and asymptotic fields \cite{sb}. However, in the
framework of \emph{exact} general relativity, progress has been
slow. In the seventies, Geroch, Hansen and others \cite{gho}
restricted themselves to the stationary context and introduced
field multipoles by analyzing the asymptotic structure of suitable
geometric fields constructed from the metric and equations they
satisfy near spatial infinity. As in electrodynamics  ---and, in
contrast to the situation in the Newtonian theory--- they found
that there are \emph{two} sets of multipoles, the mass multipoles
$\M_{(n)}$ and the angular momentum multipoles $\J_{(n)}$. In
static situations, all the angular momentum multipoles $\J_{(n)}$
vanish and the mass multipoles $\M_{(n)}$ are constructed from the
norm of the static Killing field which, like the Newtonian
gravitational potential, satisfies a Laplace-type equation outside
sources. In the (genuinely) stationary context, the $\J_{(n)}$ are
non-zero and are analogous to the magnetic multipoles in the
Maxwell theory. In the Newtonian theory, since the field
multipoles are defined as coefficients in the $1/r$ expansion of
the gravitational potential, knowing all multipoles one can
trivially reconstruct the potential outside sources. In general
relativity, there are considerable coordinate ambiguities in
performing asymptotic expansions of the metric. Therefore, Geroch
and Hansen were led to define their multipoles using other
techniques. Nonetheless, Beig and Simon \cite{bs} have established
that the knowledge of these multipoles does suffice to determine
the space-time geometry near infinity. Their construction also
shows that, if two stationary space-times have the same
multipoles, they are isometric in a neighborhood of infinity.

In the Geroch-Hansen framework, one works on the 3-dimensional
manifold of orbits of the stationary Killing field and multipole
moments $\M_{(n)}$ and the $\J_{(n)}$ arise as symmetric,
trace-free tensors in the tangent space of the point $\Lambda$ at
spatial infinity of this manifold. Now, the vector space of n-th
rank, symmetric traceless tensors on $\R^3$ is naturally
isomorphic with the vector space spanned by the linear
combinations $\sum_{m=-n}^{m=n}\, a_{nm} Y_{n}^{m}(\theta, \phi)$
of spherical harmonics on the unit sphere $\S^2$ in $\R^3$.
Therefore, each $\M_{(n)}$ and $\J_{(n)}$ uniquely defines a set
of numbers, $M_{nm}$ and $J_{nm}$ with $m \in \{-n,-n+1,\ldots,
n-1, n\}$. Finally, let us consider stationary space-times which
are also axi-symmetric (e.g., the Kerr space-time). Then $M_{n,m}
= M_n \delta_{m,0}$ and $J_{nm} = J_n \delta_{m,0}$, whence
multipoles are completely characterized by two sets of numbers,
$M_n, J_n$ with $n =0,1,\ldots$. This fact will be useful to us
because most of our analysis will be restricted to axi-symmetric
isolated horizons.

In stationary space-times, then, the situation with field
multipoles is quite analogous to that in the Newtonian theory. The
status of source multipoles, on the other hand, has not been so
satisfactory. Dixon developed a framework to define source
multipoles \cite{gd} but the program did not reach the degree of
maturity enjoyed by the field multipoles. On the other hand, it is
the source multipoles, rather than the field multipoles defined at
infinity, that appear to be more directly relevant to the motion
of extended bodies in general relativity. (For recent discussions,
see \cite{recent}.) In particular, it would be interesting to
obtain useful generalizations of the quadrupole formula to fully
relativistic objects such as neutron stars and black holes. Is
there, for example, a relation between the flux of energy falling
across the horizon and the rate of change of the quadrupole moment
of a black hole? To analyze such issues one needs a reliable
definition of the source quadrupole moment, and more generally, of
source multipoles.

In this paper, we will focus on the problem of defining the
analogs of source multipoles for black holes in equilibrium. The
isolated horizon framework provides a suitable quasi-local arena
for describing such black holes \cite{prl,abf,afk,abl1,abl2}.
Thus, our task is to introduce an appropriate definition of
multipoles which capture distortions of the horizon geometry and
of the distribution of angular momentum currents on the horizon
itself, and explore properties of these multipoles. Do these
multipoles provide a diffeomorphism invariant characterization of
the horizon geometry? If so, they may play a significant role in
the analysis of equations of motion. Do they suffice to determine
space-time geometry in the neighborhood of the horizon in
stationary space-times? These are attractive possibilities. But
since the horizon lies in a genuinely strong field region, a
priori it is not obvious that a useful notion of multipoles can be
introduced at all. Indeed, since the problem of defining and
analyzing multipoles has turned out to be so difficult already for
relativistic fluids, at first glance it may seem hopelessly
difficult for black holes. However, a tremendous simplification
occurs because black holes are purely geometric objects; one does
not have to resolve the messy and difficult issues related to the
details of matter sources. We will see that this simplification
makes it possible to carry out a detailed analysis and
satisfactorily address the issues raised above.

We will restrict our detailed analysis to axi-symmetric (or type
II) isolated horizons, where the symmetry restriction applies only
to the horizon geometry (and to the pull-back of the Maxwell field
to the horizon) and not to the entire space-time. The material is
organized as follows.  In section \ref{s2}, we recall the relevant
facts about isolated horizons and axi-symmetric geometric
structures. In section \ref{s3} we define the multipoles $M_n$ and
$J_n$ and show that these two sets of numbers provide a complete
characterization of the isolated horizon geometry. For the
specific case of Maxwell sources, we introduce another pair, $Q_n,
P_n$ of electromagnetic multipoles and show that they suffice to
determine the pull-back of the Maxwell tensor as well as that of
its dual. Finally, the initial value formulation based on two
intersecting null surfaces \cite{frl} implies that, if there is a
stationary Killing field in the neighborhood of the horizon,
multipoles also suffice to determine the near horizon geometry
\cite{abl3} . In section \ref{s4}, we discuss potential extensions
and applications of the framework ranging from numerical
relativity to quantum gravity.

Unless otherwise stated, in this paper all manifolds and fields
will be assumed to be smooth.

\section{Preliminaries}
\label{s2}

\subsection{Isolated horizons}
\label{s2.1}

In this sub-section we briefly recall the relevant notions
pertaining to isolated horizons \cite{prl,afk,abl1,abl2}. This
discussion will also serve to fix our notation.

Let us begin with the basic definitions \cite{afk}. A
\emph{non-expanding horizon} $\IH$ is a null, 3-dimensional
sub-manifold of the 4-dimensional space-time $(\man,
g_{ab})$, with topology $\S^2\times \R$, such that:\\
i) the expansion $\theta_\l$ of its null normal $\l$ vanishes;
and,\\
ii) field equations hold on $\IH$ with stress energy, $T_{ab}$,
satisfying the very weak requirement that $-T^a{}_{b}\l^b$ is a
future-directed, causal vector. (Throughout, $\l^a$ will be
assumed to be future pointing.)

Before discussing their consequences, let us note two facts about
these assumptions: i) if the expansion vanishes for one null
normal, it vanishes for all; and, ii) the condition on the stress
energy is satisfied by all the standard matter fields {provided
they are minimally coupled to gravity}. (Non-minimally coupled
matter can be incorporated by a small modification of this
condition. See, e.g., \cite{acs}.)
Since $\IH$ is a null 3-surface, its intrinsic `metric' $q_{ab}$
has signature 0,+,+. The definition ensures that 
$T^a{}_b \l^b \propto \l^a$ and ${\cal L}_\l q_{ab}
\=0$, \emph{where $\=$ denotes equality restricted to points of
$\Delta$}. In particular, the area of any 2-sphere cross-section
is constant on $\IH$. We will denote it by $a_\IH$.
The definition also implies that the space-time derivative
operator $\nabla$ naturally induces a unique derivative operator
$\D$ on $\IH$.  Furthermore, $D_a l^b = \omega_a l^b$
for some globally defined 1-form $\omega_a$ on the horizon. This
1-form will play an important role.

The pair $(q_{ab}, \D)$ is referred to as the intrinsic
\emph{geometry} of $\IH$. The notion that the black hole itself is
in equilibrium is captured by requiring that this geometry is time
independent: \\
An \emph{isolated horizon} $(\IH, [\l ])$ is a non-expanding
horizon $\IH$ equipped with an equivalence class of null normals
$\l^a$, where $\l^a \sim {\l^\prime}^a$ if and only if
${\l^\prime}^a = c \l^a$ for a positive constant $c$, such that
$[{\cal L}_\l, \, D] \= 0$ on $\IH$.%
\footnote{Given $(\Delta, q_{ab}, \D)$, one can show that,
generically, there is an unique equivalence class $[\l]$ of null
normals such that $(\Delta, [\ell])$ is an isolated horizon.
However, our analysis will not be restricted to this case.}

Since $\Delta$ is a null surface, given any one null normal $\l^a$
in $[\l]$, we have $\l^a D_a\l^b = \k_{\l} \l^b$ for some $\k_\l$.
(Thus, $\k_\l = \l^a \omega_a$.) The requirement $[{\cal L}_\l,\,
\D] \= 0$ further implies that $\k_\l$ is constant on $\IH$.  It
is referred to as the \emph{surface gravity} of $\IH$ with respect
to $\ell^a$. Note that if ${\l^\prime}^a = c\l^a$ then
$\k_{\l^\prime} = c\k_\l$. Thus, the \emph{value} of surface
gravity refers to a specific null normal; it is not a property of
$(\IH, [\l])$. However, one \emph{can} unambiguously say whether
the given isolated horizon is \emph{extremal} (i.e., has $\k_\l
=0$) or \emph{non-extremal} (i.e., has $\k_\l \not= 0$).

The isolated horizon definition extracts from the notion of the
Killing horizon just that `tiny' part which turns out to be
essential for black hole mechanics \cite{afk,abl2} and, more
generally, to capture the notion that the horizon is in
equilibrium, allowing for dynamical processes and radiation in the
exterior region \cite{prl}. Indeed, Einstein's equations admit
solutions with isolated horizons in which there is radiation
arbitrarily close to the horizons \cite{pc}. Finally, note that
the definition uses conditions which are \emph{local} to $\IH$.
Hence, unlike event horizons one does not require the knowledge of
full space-time; the notion is not `teleological'.

Of particular interest to our analysis is the case when the only
matter present \emph{at} the horizon is a Maxwell field. In this
case, only the \emph{pull-backs} $B_{ab}$ and $E_{ab}$ to $\IH$,
respectively of the Maxwell field $F_{ab}$ and its dual
${\dual{F}_{ab}}$ are needed in the isolated horizon analysis.
\emph{We will refer to the quadruplet $(q_{ab}, \D, B_{ab},
E_{ab})$ as the Einstein-Maxwell geometry of the isolated
horizon.}

\textsl{Remark:} Note that the notion of an isolated horizon
$(\IH, [\l])$, can be formulated intrinsically, using only those
fields which define its geometry, without reference to the full
space-time metric or curvature. Specifically: i) the condition
$\l^a$ is a null normal to $\IH$ is captured in the property that
$q_{ab}$ has signature 0,+,+ with $q_{ab}\l^b \=0$;  ii) the
condition $\theta_\l \=0$ can be replaced by $\Lie_\l q_{ab} \=0$;
iii) the requirement on the stress-energy refers only to fields
$B_{ab}$ and $E_{ab}$; and iv) the condition $[\Lie_\ell, \, \D]
\= 0$ refers only to $\D$ (and not to the full space-time
connection). \emph{This is why the quadruplet $(q_{ab}, \D,
B_{ab}, E_{ab})$, defined intrinsically on $\IH$, was singled out
to introduce the notion of Einstein-Maxwell horizon geometry.} All
equations that are required in the derivation of the laws of
mechanics of isolated horizons in Einstein-Maxwell theory
\cite{afk,abl2} as well as the main results of this paper refer just
to these fields.

Next, let us examine symmetry groups of isolated horizons. A
\emph{symmetry} of $(\IH, \l^a, q_{ab}, \D, B_{ab}, E_{ab})$ is a
diffeomorphism on $\IH$ which preserves $q_{ab}, \D, B_{ab}$ and
$E_{ab}$ and at most rescales $\l^a$ by a positive constant. It is
clear that diffeomorphisms generated by $\l^a$ are symmetries. So,
the symmetry group $G_\IH$ is at least 1-dimensional. The question
is: Are there any other symmetries? At infinity, we generally have
a universal symmetry group (such as the Poincar\'e or the anti-de
Sitter) because all metrics under consideration approach a fixed
metric (Minkowskian or anti-de Sitter) there. In the case of the
isolated horizons, generically we are in the strong field regime
and space-time metrics do not approach a universal metric.
Therefore, the symmetry group is not
universal. However, there are only three universality classes:\\
i) Type I: the isolated horizon geometry is spherical; in this
case, $G_\IH$ is four dimensional;\\
ii) Type II: the isolated horizon geometry is axi-symmetric; in
this case, $G_\IH$ is two dimensional;\\
iii) Type III: the diffeomorphisms generated by $\l^a$ are the
only symmetries; $G_\IH$ is one dimensional.

Note that these symmetries refer \emph{only to} the horizon
geometry. The full space-time metric need not admit any isometries
even in a neighborhood of the horizon. Physically, type II
horizons are the most interesting ones. They include the
Kerr-Newman horizons as well as their generalizations
incorporating distortions (due to exterior matter or other black
holes) and hair. Our main results refer to the type II case.

Finally, one can use the field equations to isolate the
\emph{freely specifiable data} which determines the
Einstein-Maxwell geometry of an isolated horizon \cite{abl1, lp}.
The analysis is naturally divided in to two cases: $\k_\l \not= 0$
and $\k_\l = 0$. Fix any 2-sphere cross-section $\t\IH$ of $\IH$
and denote by $\t{q}^a{}_b$ the natural projection operator on
$\t\IH$. Then, on any type II horizon, we have the following.%
\footnote{For convenience of the reader, proofs of these
assertions are sketched in Appendix A. Throughout this paper, the
term `free data' refers to freely specifiable \emph{fields}. In
order to reconstruct the horizon geometry, in addition to
specifying this data one also needs to fix some constants, such as
the horizon area $a_\IH$.}
\begin{itemize}
\item  In the \emph{non-extremal case}, the free data consists of
the projections $\tq_{ab}, \tw_a, \tB_{ab}, \tE_{ab}$ to $\t\IH$
(using $\t{q}^a{}_b$) of $q_{ab}, \omega_a, {B}_{ab}, {E}_{ab}$ on
$\IH$. (Recall that $\omega_a$ is defined by $\D_a \l^b = \omega_a
\l^b$.) That is, given the free data on $\t\IH$, using the
projections of the field equations on $\t\IH$, one can reconstruct
the full Einstein-Maxwell geometry $(q_{ab}, \D, B_{ab}, E_{ab})$
on $\IH$.
\item  In the \emph{extremal} case, remarkably, the fields
$q_{ab}, \omega_a, B_{ab}$ and $E_{ab}$ turn out to be all
\emph{universal}; they are the same as those in the extremal
Kerr-Newman geometry \cite{lp}! The free data consists of
$(\tS_{ab})$, a symmetric, second rank tensor on $\t\IH$,
determined by $\D$ via $\t{S}_{ab} = \t{q}_a{}^c \t{q}_b{}^d
D_c n_d$ where $n_a$ is the covariant normal to $\t\IH$ within
$\IH$, satisfying $\l^a n_a = -1$. (Thus, $\t{S}_{ab}$ is the
analog of the `extrinsic curvature of $\t\IH$ within $\IH$.) Given
a Kerr-Newman quadruplet $q_{ab}, \omega_a, B_{ab}$ and $E_{ab}$
the only freedom in the Einstein-Maxwell horizon geometry lies in
the part of $\D$ which determines the `extrinsic curvature'
$\t{S}_{ab}$ of $\t\IH$.
\end{itemize}
These results will play an important role in section 3.
Specifically, we will construct multipoles from the free data and
show that they provide a diffeomorphism invariant characterization
of the free data and hence, via field equations, of the horizon
geometry.%
\footnote{This general philosophy is rather similar to that used
by Janis and Newman to introduce a notion of multipoles using
`free data' on null surfaces \cite{jn}. However, since our
analysis is restricted to isolated horizons rather than general
null surfaces, our results are free of coordinate ambiguities.}

\subsection{Axi-symmetric structures}
\label{s2.2}

In this sub-section we will recall a few facts about axi-symmetric
geometries on $\S^2$. This discussion will also make our assumptions
explicit.

Let $\S$ be a manifold with the topology of a 2-sphere, equipped
with a metric $\t{q}_{ab}$. We will denote by $\t\epsilon_{ab}$
the alternating tensor on $\S$ compatible with $\tq_{ab}$, by $a$
the area of $\S$ and by $R$ its radius, defined by $a = 4\pi R^2$.
We will say that $(\S, \t{q}_{ab})$ is \emph{axi-symmetric} if it
admits a Killing field $\phi^a$ with closed orbits which vanishes
\emph{exactly} at two points of $\S$. The two points will be
referred to as \emph{poles}. We will now show that such metric
manifolds carry invariantly defined coordinates and discuss
properties of the metric coefficients in these coordinates.

Since $\Lie_\phi \t\epsilon_{ab} =0$, there exists a unique
(globally defined) function $\zeta$ on $\S$ such that
\be \label{zeta}\t\D_a \zeta = \f{1}{R^2}\, \t\epsilon_{ba} \phi^b
\quad {\rm and} \quad \oint_{\S} \zeta \, \t\epsilon  = 0 \, .\ee
It is clear that $\Lie_\phi \zeta \=0 $ and $\t\D_a \zeta$
vanishes \emph{only} at the poles. Hence, $\zeta$ is a monotonic
function on the 1-dimensional manifold $\hat\S$ of orbits of
$\phi^a$. Now $\h\S$ is a closed interval and the end points
correspond to the two poles. Hence $\zeta$ monotonically increases
from one pole to another. We will say that $\zeta$ assumes its
minimum value at the \emph{south pole} and the maximum at the
\emph{north pole}.

Next, let us introduce a vector field $\zeta^a$ on $\S^\prime =
(\S -\,{\rm poles})$ via:
\be \t{q}_{ab} \zeta^a \phi^b = 0 \quad {\rm and}\quad
\zeta^a\t\D_a \zeta = 1\ee
Then it follows that
\be \label{zetaa}\zeta^a = \f{R^4}{\mu^2}\,\, \tq^{ab}\, \t\D_b
\zeta\, ,\ee
where $\mu^2 = \tq_{ab} \phi^a\phi^b$ is the squared norm of
$\phi^a$. Hence integral curves of $\zeta^a$ go from the south
pole to the north (and $\zeta^a$ diverges as one approaches the
poles). Using $\zeta^a$, we can define a preferred affine
parameter $\phi$ of $\phi^a$ as follows: Fix any one integral
curve $I$ of $\zeta^a$ in $\S^\prime$ and set $\phi =0$ on $I$.
Thus, on $I$, we have $\Lie_\zeta \phi =0$. Now, since $\phi^a$ is
a Killing field and $\zeta^a$ is constructed uniquely from
$(\tq_{ab}, \phi^a)$, it follows that $\Lie_{\phi} \zeta^a =0$.
Hence, we conclude: $ \Lie_\phi (\Lie_\zeta \phi) =0$. Since
$\Lie_\zeta \phi = 0$ on $I$, it now follows that $\phi = {\rm
const}$ on \emph{every} orbit of $\zeta^a$. This now implies that
the affine parameter
$\phi$ of $\phi^a$ has the same range on every orbit of $\phi^a$.%
\footnote{To make this argument precise, one should work on the
universal covering of the orbits of $\phi^a$ on $\S^\prime$ but
the additional steps are straightforward.}
Without loss of generality we will assume that $\phi^a$ is such that
$\phi \in [0, 2\pi)$ on $\S$.

Thus, starting from geometry, we have constructed two coordinates
$\zeta, \phi$ on $\S$ such that $\phi^a \equiv
(\partial/\partial\phi)^a$ and $\zeta^a \equiv
(\partial/\partial\zeta)^a$ are orthogonal. Eqs (\ref{zeta}) and
(\ref{zetaa}) now imply that the metric has the form:
\be\label{q} \tq_{ab} = R^2(f^{-1} D_a\zeta\, D_b\zeta + f
D_a\phi\, D_b\phi) \quad {\rm and} \quad \tq^{ab} = \f{1}{R^2}(f
\zeta^a \zeta^b + f^{-1} \phi^a \phi^b) \ee
where $f = \mu^2/R^2$. The fact that the area of $\S$ is $4\pi
R^2$ now implies that the range of $\zeta$ is necessarily
$[-1,1]$. Conversely, given any `coordinates' $\zeta^\prime,
\phi^\prime$ in $[-1,1]\times \R/2\pi {\mathbb Z}$ in which the
metric can be expressed in (the primed version of) the form
(\ref{q}), it is easy to show that $\zeta^\prime = \zeta$ and
$\phi^\prime$ is at most a rigid shift of $\phi$. (For an analogous
construction in Newman-Penrose notation, see \cite{Date}.)

Functions $\zeta,\phi$ serve as `coordinates' on $\S$ modulo usual
caveats: they are ill-defined at the poles and $\phi$ has a $2\pi$
discontinuity on the integral curve $I$ of $\zeta^a$. We have to
ensure that the metric $\tq_{ab}$ is smooth in spite of these
coordinate problems. The discontinuity at $I$ causes no problems.
However, poles do require a careful treatment because the norm
$\mu$ of $\phi^a$ ---and hence $f$--- vanishes there. Smoothness
of $\tq_{ab}$ at the poles (i.e. absence of conical singularities)
imposes a non-trivial condition on $f$:
\be \label{bc} \lim_{\zeta\to \pm 1}\, f'(\zeta) = \mp 2 \ee
where `prime' denotes derivative with respect to $\zeta$. On a
metric 2-sphere, we have $f = 1 -\zeta^2$ and we can bring the
metric to the standard form simply by setting $\zeta = \cos\theta$.
In the general case, $f$ has the same values and first derivatives
at the poles as on a metric 2-sphere. Using l'Hopital's rule, one
can show that this fact suffices to ensure that the metric
(\ref{q}) is smooth at the poles.

Finally, we note a property of these axi-symmetric metrics which
will be useful in section \ref{s3.2}. A simple calculation shows
that the scalar curvature $\tR$ of $\tq_{ab}$ is given by:
 \be \label{tR} \tR (\zeta,\phi) = - \f{1}{R^2}\, f''(\zeta)\, . \ee
By integrating it twice with respect to $\zeta$ \emph{and using
the boundary conditions} $(f|_{\zeta= -1}) = 0$ and $(f'|_{\zeta =
-1}) = 2$, one can reconstruct the function $f$ from the scalar
curvature:
\be f = - R^2 \left[\int_{-1}^\zeta \d\zeta_1\,
\int_{-1}^{\zeta_1} \d \zeta_2 \,\tR(\zeta_2)\right] \,\,\, + \,
 2(\zeta +1) \ee
Thus, thanks to the preferred coordinates admitted by an
axi-symmetric geometry on $\S$, given the area $a$ of $\S$ and the
scalar curvature $\tR$, the metric $\tq_{ab}$ is completely
determined.

\emph{Remark:} Coordinates  $(\zeta,\phi)$, determined by the
axi-symmetry of $\tq_{ab}$, also enable us to define a
\emph{canonical} round, 2-sphere metric $\tq^o_{ab}$ on $\S$:
\be \label{qo} \tq^o_{ab} = R^2(f_0^{-1} D_a\zeta\, D_b\zeta + f_o
D_a\phi\, D_b\phi) \ee
where $f_o = 1-\zeta^2$. Note that $\tq^o_{ab}$ has the \emph{same
area element} as $\tq_{ab}$. This round metric captures the extra
structure made available by axi-symmetry in a coordinate invariant
way. The availability of $\tq^o_{ab}$ enables one to perform a
natural spherical harmonic decomposition on $\S$. This fact will
play a key role throughout section \ref{s3}.

\section{Multipoles of type II isolated horizons}
\label{s3}

This section is divided into four parts. In the first three, we
restrict ourselves to non-extremal, type II isolated horizons with
no matter fields on them. We begin in the first part by defining a
set of multipoles, $I_n, L_n$, starting from the horizon geometry.
We then show that one can reconstruct the horizon geometry
starting from these two sets of numbers. In the second, we show that
if two isolated horizons have the same multipoles $I_n, L_n$,
their geometries are related by a diffeomorphism. Thus, these two
sets of numbers provide a convenient diffeomorphism invariant
characterization of the horizon geometry. Therefore, we refer to
$I_n, L_n$ as \emph{geometric multipoles}. In the third part, we
rescale these moments by appropriate dimensionful factors to
obtain mass and angular momentum multipoles $M_n, J_n$. Finally,
in the fourth part we first discuss electromagnetic multipoles and
then summarize the situation on extremal isolated horizons.

\subsection{Geometric Multipoles}
\label{s3.1}

Let $(\IH, [\l])$ be a non-extremal, type II isolated horizon with
an axial Killing field $\phi^a$. In this sub-section, we will
ignore matter fields on $\IH$ and concentrate just on the horizon
geometry defined by $(q_{ab}, \D)$. Fix a cross-section $\t\IH$ of
$\IH$. Then, as summarized in section \ref{s2.1}, the free data
that determine the horizon geometry consists of the pair
$(\tq_{ab}, \tw_a)$ where $\tq_{ab}$ is the intrinsic metric on
$\t\IH$ and $\tw_a$ is the projection on $\t\IH$ of the 1-form
$\omega_a$ on $\IH$ (defined by $\D_a \l^b = \omega_a \l^b$).
However, there is some gauge freedom associated with our choice of
the cross-section $\t\IH$ \cite{abl1}. We will first spell it out
and then define multipoles using gauge invariant fields.

For simplicity of presentation, let us fix a null normal $\l^a$ in
$[\l^a]$. Then, $\t\IH$ can be regarded as a leaf of a foliation
$u = {\rm const}$ such that $\l^a \D_a u \= 1$. For notational
simplicity, we will set $n_a =-D_a u$ so that $n_a$ is the
covariant normal to the foliation satisfying $\l^a n_a = -1$. The
projection operator $\t{q}_a{}^b$ on the leaves of this foliation
is given by $\t{q}_a{}^b = \delta_a^b + n_a \l^b$. Hence,
$\tq_{ab} = q_{ab}$ and $\tw_a = \omega_a + \k_\l n_a$ as tensor
fields on $\IH$. Since $\Lie_\l q_{ab} \=0$ and $\Lie_\l \omega_a
\= 0$ on any isolated horizon, and since $\Lie_\l n_a \=0$ from the
definition of $n_a$, it follows that $(\tq_{ab}, \tw_a)$ on any
one leaf is mapped to that on any other leaf under the natural
diffeomorphism (generated by $\l^a$) relating them. Let us now
consider a cross-section $\t\IH^\prime$ which does \emph{not}
belong to this foliation. Let $u^\prime = {\rm const}$ denote the
corresponding foliation. Set $F = u- u^\prime$. Then, regarded as
tensor fields on $\IH$ the two sets of free data are related by
\be \tq^\prime_{ab} = \tq_{ab} \quad {\rm and} \quad \tw^\prime_a
= \tw_a + \k_\l\, \D_a F \ee
Thus, under the natural diffeomorphism (defined by the integral
curves of $\l^a$) between $\t\IH$ and $\t\IH^\prime$, $\tq_{ab}$
is mapped to $\tq_{ab}^\prime$ but $\tw_a$ is \emph{not} mapped to
$\tw_a^\prime$; the difference is a gradient of a function. This
is the gauge freedom in the free data.

It is therefore natural to consider, in place of $\tw_a$, its
curl. From the isolated horizon framework, it is known that for
any null tetrad $\l^a, n^a, m^a, \bar{m}^a$ such that $\ell^a \in
[\ell]$, the Weyl components $\Psi_0$ and $\Psi_1$ vanish on $\IH$
whence $\Psi_2$ is gauge invariant \cite{afk, abl1}, and the curl
of $\tw_a$ is given just by $\Im \Psi_2$:
\be \label{Psi2}  \D_{[a}\t\omega_{b]} = \Im \Psi_2
\,\epsilon_{ab} \ee
where $\epsilon_{ab}$ is the natural area element on $\IH$ (which
satisfies $\epsilon_{ab}\l^a \=0$ and $\Lie_{\l} \epsilon_{ab} \=
0$). Thus, the gauge invariant content of $\t\omega_a$ is coded in
$\Im \Psi_2$.

The second piece of free data is the metric $\tq_{ab}$ on $\t\IH$.
In section \ref{s2.2} we showed that using an invariant coordinate
system $(\zeta, \phi)$, one can completely determine $\tq_{ab}$ in
terms of a number, the area $a$ and a function, its scalar
curvature $\tR$. If, as assumed in this sub-section, the
cosmological constant is zero and there are no matter fields on
$\IH$, then $\tR = - 4\Re \Psi_2$ \cite{abf}. Hence the gauge
invariant part of the free data that determines the horizon
geometry is neatly coded in the Weyl component $\Psi_2$.

It is therefore natural to define multipoles using a complex
function $\Phi_\IH$ on $\Delta$:
\be \label{Phi} \Phi_\IH := \f{1}{4} \tR - i \Im \Psi_2 \, . \ee
(Thus, in absence of matter on $\IH$, $\Phi_\IH = -\Psi_2$ while
in presence of matter it is given by $\Phi_\IH \= - \Psi_2 +
({1}/{4}) R_{ab} \tq^{ab} - ({1}/{12}) R$, where $R_{ab}$ is the
Ricci tensor and $R$ the scalar curvature of the 4-metric at the
horizon.) Since all fields are axi-symmetric, using the natural
coordinate $\zeta$ on $\t\IH$, \emph{we are led to define
multipoles as}:
\be \label{def} I_n + i L_n := \oint_{\t\IH}\, \Phi_\IH\,
Y^0_n(\zeta)\,\, \d^2\t{V} \ee
or,
$$ I_n := \f{1}{4}\, \oint_{\t\IH} \tR\, Y^0_n(\zeta)\,
\d^2\t{V}\quad {\rm and} \quad L_n := - \oint_{\t\IH}\,
\Im\Psi_2\, Y^0_n(\zeta)\,\, \d^2\t{V}. $$
Here $Y^0_n$ are the $m=0$ spherical harmonics, subject to the
standard normalization:
\be \oint_{\t\IH} Y_n^0 Y_m^0 \,\d^2\t{V} = R^2_\IH \delta_{n,m}\,
, \ee
where $R_\IH$ is the horizon radius defined through its area
$a_\IH$ via $a_\IH = 4\pi R^2_\IH$. Thus, given any horizon
geometry, we can define a set of two numbers, $I_n$ and $L_n$.
Recall that $\zeta$ is determined entirely by the metric
$\tq_{ab}$ and the rotational Killing field $\phi^a$. Therefore,
it is immediate that if $(\IH, q_{ab}, \D)$ and $(\IH^\prime,
q^\prime_{ab}, \D^\prime)$ are related by a diffeomorphism, we
have $I_n = I_n^\prime$ and $L_n = L_n^\prime$; the two sets of
numbers are diffeomorphism invariant. If the isolated horizon were
of type I, $q_{ab}$ would be spherically symmetric and $\Im
\Psi_2$ would vanish \cite{afk}. Then, the only non-zero multipole
would be $I_0$ which, by the Gauss-Bonnet theorem, has a universal
value (see below). Given a generic type II horizon, as we saw in
section \ref{s2.2}, the axi-symmetric structure provides a
canonical 2-sphere metric $q^o_{ab}$ (which, in the type I case,
coincides with the physical metric). The physical geometry has
distortion and rotation built in it. The round metric $q^o_{ab}$
serves as an invariantly defined `background' against which one can
measure distortions and rotations. Multipoles $I_n, L_n$ provide a
diffeomorphism invariant characterization of these. More
precisely, they encode the difference between the physical horizon
geometry $(q_{ab}, \D)$ and the fiducial, type I geometry
determined by $(\tq^o_{ab}, \tw_a =0)$.

Finally, note that $I_n$ and $L_n$ can not be specified entirely
freely but are subject to certain algebraic constraints. The first
comes from the Gauss-Bonnet theorem which, in the axi-symmetric
case, follows from the boundary condition (\ref{bc}) on $f'$ and
the expression (\ref{tR}) of the scalar curvature in terms of $f$:
\be \label{I0} I_0 = \f{1}{4}\, \oint_{\t\IH}\, \tR\,
Y_0^0(\zeta)\, \d^2\t{V} = \sqrt{\pi}\, .\ee
The second comes directly from the relation (\ref{Psi2}) between
$\Im\Psi_2$ and curl of $\tw_a$:
\be \label{L0} L_0 = -\, \f{1}{\sqrt{4\pi}}\oint_{\t\IH}\,
\Im\Psi_2\,  \d^2\t{V} = 0\ee
The third constraint comes again from (\ref{bc}) and (\ref{tR}):
\be \label{I1} I_1 := \f{\sqrt{3}}{8\sqrt{\pi}}\, \oint_{\t\IH}\,
\tR \,\zeta\, \d^2\t{V} = 0\ee
We will show in section \ref{s3.3} that (\ref{L0}) implies that,
as one would physically expect, the `angular momentum monopole'
necessarily vanishes and (\ref{I1}) implies that the mass dipole
vanishes, i.e., that our framework has automatically placed us in
the `center of mass frame of the horizon'. Next, because
$\Phi_\IH$ is smooth, these moments have a certain fall-off. Let
us assume that $\Phi_\IH$ is $C^k$ (i.e., the space-time metric is
$C^{k+2}$). Then as $n$ tends to infinity, $I_n$ and $L_n$ must
fall off in such a way that
\be\label{falloff} \sum_{n=0}^\infty\, \sum_{m=0}^{k+1} |n^{m}
(I_n - iL_n)|^2 < \infty \ee
Finally, there is a constraint arising from the fact that $f$ is
non-negative and vanishes only at the poles. (This property of $f$
is essential for regularity of the metric.) Using (\ref{tR}) and
the definition (\ref{def}) of $I_n$, one can express $f$ in terms
of $I_n$. Unfortunately, the resulting restriction on multipoles
is quite complicated:
\ba \label{posf} f(\zeta) = 1 - \zeta^2 &+&
\sum_{n=2}^{\infty}\,\, \f{2}{\sqrt{\pi(2n+1)}}\,\,\big[ -
\f{1}{2n+3}\, P_{n+2}(\zeta) \nonumber\\
&+& \f{2(2n+1)}{(2n+3)(2n-1)}\, P_n (\zeta) - \f{1}{2n-1}\,
P_{n-2}(\zeta)\big]\, I_n\,\, \ge 0 \ea
and vanishes only if $\zeta =\pm 1$, where $P_n$ are the Legendre
polynomials.

\textsl{Remark:} We conclude by noting some simplifications that
occur in presence of additional symmetries. Certain space-time
metrics, such as the Kerr solutions have a \emph{discrete
(spatial) reflection symmetry}, $(\zeta, \phi) \mapsto (-\zeta,
\phi+\pi)$, under which $\Psi_2 \mapsto \Psi_2^\star$. Therefore,
in the isolated horizon framework it is interesting to consider
the case in which $\Phi_\Delta \mapsto \Phi_\Delta^\star$ and
$\epsilon_{ab} \mapsto \epsilon_{ab}$ under the discrete
diffeomorphism $\zeta \mapsto -\zeta$ on $\IH$. Then, since
$Y_n^0$ are even/odd under reflections if $n$ is even/odd, it
follows that $I_n =0$ for all odd $n$ and $L_n =0$ for all even
$n$. Next, consider the case in which the isolated horizon is a
Killing horizon of a \emph{static} Killing field. Then, one can
show that $\Im \Psi_2 \=0$ \cite{abl2}. Hence $L_n =0$ for all $n$
but in general $I_n$ can be arbitrary, capturing possible
distortions in the horizon geometry. Finally, consider \emph{type
I isolated horizons} on which the horizon geometry is spherically
symmetric. Then, $\Phi_\Delta$ and $\epsilon_{ab}$ are spherically
symmetric. It is then obvious from properties of spherical
harmonics $Y_n^0(\zeta)$ that for all $n>0$, we have $I_n = L_n
=0$. Since $L_0$ always vanishes, in this case the only
non-trivial multipole is $I_0 = \sqrt{\pi}$.

\subsection{Reconstruction of the horizon geometry and the uniqueness issue}
 \label{s3.2}

\subsubsection{Construction}
\label{s3.2.1}

Let us first show that the knowledge of the area and multipole
moments suffices to reconstruct an isolated horizon geometry.

\emph{ Suppose we are given only the radius $R_\IH$ and the set
$\{I_n, L_n\}$ of geometric multipoles of
a $C^k$, non-extremal, type II isolated horizon. Then, one can
explicitly construct a $C^k$, non-extremal, type II isolated
horizon geometry $(\IH, q_{ab}, \D)$ such that area is given by
$a_\IH = 4\pi R^2_\IH$, and geometric multipoles are given by
$I_n$ and $L_n$.}

Let $\S$ be a smooth 2-manifold, topologically $\S^2$. Let
$(\zeta, \phi)$ be a spherical coordinate system on $\S$, with
$\zeta \in [-1,1]$ and $\phi \in [0, 2\pi)$. Define
\ba \label{tr} \tR &:=& \frac{4}{R^2_\IH}\, \sum_{n=0}^\infty
I_n\, Y_n^0(\zeta)
\\
\label{impsi2} \Im{\Psi_2}&:=& -\, \frac{1}{R^2_\IH}\,
\sum_{n=0}^\infty L_n Y_n^0(\zeta). \ea
Since $\{I_n,L_n\}$ are multipoles of a $C^k$ isolated horizon, we
are guaranteed that the fields $\tR$ and $\Im \Psi_2$ so defined
are $C^k$. Next, define a function $f$ on $\S$ via:
\be f(\zeta):= -\left[R^2_\IH\, \int_{-1}^\zeta \d \zeta_1
\int_{-1}^{\zeta_1} \d \zeta_2 \, \tR(\zeta_2)\right]\,\,
+\,2(\zeta+1) \ee
and set
\be {\tq}_{ab}:= R^2_\IH \left( \frac{1}{f(\zeta)}\, \D_a \zeta
\D_b \zeta + f(\zeta)\, D_a\phi\D_b \phi \right). \ee
Since $I_n$ are the multipoles of an isolated horizon,
(\ref{posf}) is satisfied whence it follows that $f$ is
non-negative on $\S$ and vanishes only at the `poles', $\zeta =
\pm 1$. Hence $q_{ab}$ is a smooth metric except possibly at the
poles. Next, $I_0=\sqrt{\pi}$ and $I_1 =0$. Interestingly, these
conditions imply $f'(\pm 1)= \mp 2$, and therefore ensure that the
metric $\tq_{ab}$ can be smoothly extended to points $\zeta=\pm 1$
of $\S$. Next, it is straightforward to verify that the area of
$\S$ with respect to this metric is given by $4\pi R^2_\IH$ and
its scalar curvature is given by $\tR$. Let us now turn to the
multipoles $L_n$. Since these multipoles also come from a type II
isolated horizon geometry, we have $L_0 =0$. Eq (\ref{impsi2}) now
implies $\oint \Im \Psi_2\, \d^2\t{V} = 0$, whence there is a
globally defined 1-form $\tw_a$ on $\S$ such that
\be \D_{[a} \tw_{b]} = \Im \Psi_2\,\, \epsilon_{ab} \ee
where $\epsilon_{ab}$ is the alternating tensor on $(\S,
\tq_{ab})$.

Now consider a 3-manifold $\IH = \S\times \R$ and equip it with a
vector field $\l^a$ along the `$\R$-direction'. Let $\pi$ be an
embedding of $\S$ onto a 2-sphere cross-section $\t\IH$ of $\IH$.
Then, the diffeomorphism $\pi$ equips $\t\IH$ with fields
$\tq_{ab}, \tw_a$. This is the free data required to construct a
non-extremal isolated horizon geometry $(q_{ab}, \D)$ on $\IH$
(see \cite{abl1} or Appendix A). Since the data are axi-symmetric,
so is the horizon geometry. Finally, from our definitions of
multipole moments, it follows immediately that the multipole
moments of this isolated horizon are given by $\{I_n, L_n\}$.

\emph{Remark:} Our construction in fact suffices to show the
following stronger \emph{existence result}: Given a positive
number $R_\IH$ and a set of real numbers $I_n, L_n$ subject to the
constraints Eqs (3.6)-(3.10), there is a non-extremal isolated
horizon geometry $(\IH, q_{ab}, \D)$ with area $a_\IH = 4\pi
R_\IH^2$ and geometric multipoles $I_n,L_n$. We chose not to refer
to the conditions (3.6)-(3.10) but assumed instead that the $I_n,
L_n$ arise from a type II isolated horizon for simplicity.
Although the existence result \emph{is} stronger, it is difficult
to verify if a given set $(I_n, L_n)$ satisfies Eqs (3.9) and
especially (3.10). The overall situation is similar at spatial
infinity. There, the constraints that must be imposed on field
multipoles to ensure that the formal power series converges to a
non-degenerate metric are awkward to state and investigations have
focussed on reconstruction \cite{gho}.

\subsubsection{Uniqueness} \label{s3.2.2}

The question now is whether the geometry we constructed in section
\ref{s3.2.1} is diffeomorphic to the one from which the multipoles
were first constructed.

\emph{Let $(\IH, [\l], q_{ab}, \D )$ and $(\IH^\prime,
[\l^\prime], q^\prime_{ab}, \D^\prime)$ be two type II
non-extremal isolated horizons without matter fields, with same
area and same geometric multipoles $I_n, L_n$. Then there is a
diffeomorphism from $\IH^\prime$ to $\IH$ which maps
$([\l^\prime], q^\prime_{ab}, \D^\prime)$ to $([\l], q_{ab}, D)$.}

Choose null normals $\l \in [\l]$ and $\l^\prime \in [\l^\prime]$
such that $\k_\l = \k_{\l'} = \k$ and consider foliations $u =
{\rm const}$ and $u' = {\rm const}$ of $\IH$ and $\IH'$ which are
tangential to $\phi^a$ and $\phi^{\prime a}$ respectively such
that $\l^a \D_a u = 1$ and $\l^{\prime a}\D_a u' =1$. The
intrinsic metrics on these leaves are axi-symmetric. Hence, as in
section \ref{s2.2}, we can find invariant coordinates $(\zeta,
\phi)$ and $(\zeta', \phi')$ on the two sets of cross-sections.
Let $\pi$ be the diffeomorphism from $\IH$ to $\IH'$ defined by
$(u, \zeta, \phi) \mapsto (u',\zeta', \phi')$. Since the
multipoles $I_n$ are the same, it follows from (\ref{tr}) that
$\pi$ maps $\tR'$ to $\tR$ and hence $\tq'_{ab}$ to $\tq_{ab}$.
Similarly, since the multipoles $L_n$ are the same, it follows
from (\ref{impsi2}) that $\pi$ maps $\Im \Psi_2'$ on $\IH'$ to
$\Im \Psi_2$ on $\IH$. But this only implies that $\d\tw'$ is
mapped to $\d\tw$, i.e.
\be \pi^*\, \tw'_a = \tw_a + \k \D_a h \ee
for some function $h$. Thus, in general, $\pi$ does not map the
free data on $\IH'$ to that on $\IH$. However, as the discussion
in the beginning of sub-section \ref{s3.1} suggests, this can be
easily remedied by changing the foliation on $\IH$. Define
$\bar{u} = u - h$ on $\IH$. Then, the projection $\bar{\omega}_a$
of $\omega_a$ on the new foliation is given by $\bar\omega_a =
\tw_a + \k \D_a h$. Hence the diffeomorphism $\bar\pi: (\bar u,
\zeta, \phi) \mapsto (u', \zeta', \phi')$ from $\IH$ to $\IH'$
does map the primed free data to the barred free data:
\be \bar\pi^*\, \tw'_a = \bar\omega_a  \quad {\rm and} \quad
\bar\pi^*\, \tq'_{ab} = \bar{q}_{ab} \equiv \tq_{ab} \ee
This isomorphism between the two sets of free data naturally
extends to an isomorphism between the two horizon geometries (see
\cite{abl1} or Appendix A).

Thus, together with area, multipoles $I_n, L_n$ provide a
convenient, diffeomorphism invariant characterization of type II
horizon geometries.

\subsection{Mass and angular Momentum multipoles}
\label{s3.3}

As is obvious from their definition, $I_n, L_n$ are all
dimensionless. Therefore, it is difficult to attribute a direct
physical interpretation to them. In this sub-section we will argue
that they can be rescaled in a natural fashion to obtain
quantities which can be interpreted as mass and angular momentum
multipoles $M_n$ and $J_n$.

In the isolated horizon framework, the area $a_\IH$ is defined
geometrically. One then defines the horizon angular momentum
$J_\IH$ as a surface term in the expression of the generator of
rotations, evaluated on a 2-sphere cross-section of the horizon.
$J_\IH$ is unambiguous because type II horizons come with an axial
symmetry \cite{abl2}. The horizon mass $M_\IH$ is also defined
using Hamiltonian methods as the generator of a preferred time
translation. However, the preferred time translation varies from
space-time to space-time. If $J_\IH = 0$, it points along $\l^a$;
if not, it is a suitable linear combination of $\l^a$ and
$\phi^a$, which can be fixed only \emph{after} one knows the value
of $J_\IH$ \cite{abl2}. Thus, $J_\IH$ is defined first before one
can fix $M_\IH$. In the same spirit, we will first define the
angular momentum multipoles $J_n$ and \emph{then} the mass
multipoles $M_n$.

We begin by recalling a general fact about angular momentum. Fix a
space-time $(\man, g_{ab})$ and a space-like 2-sphere $S$ in it.
Let $\varphi$ be any vector field tangential to $S$. Then, by
regarding $S$ as the inner boundary of a partial Cauchy surface
$M$, one can use the Hamiltonian framework to define a `conserved'
quantity $J^\varphi_{S}$
\be J^\varphi_{S}\, = - \f{1}{8\pi G}\, \oint_{S} K_{ab}\varphi^a
dS^b \ee
where $K_{ab}$ is the extrinsic curvature of $M$. In a general
space-time, this quantity is independent of $M$ if and only if
$\varphi$ is divergence free with respect to the natural area
element of $S$. Thus, for each divergence-free $\varphi$ on $S$,
$J^\varphi_{S}$ depends only on $S$ and can be interpreted as the
$\varphi^a$-component of a `generalized angular momentum'
associated with $S$. If $S$ happens to be a cross-section of
$\IH$, as one would expect, one can recast this expression in terms of
the fields defined by the isolated horizon geometry, making no
reference at all to the partial Cauchy surface $M$ \cite{abl2}:
\be \label{J} J^\varphi_S \, \= -\f{1}{8\pi G} \oint_S \varphi^a
\t\omega_a \, d^2V \=  -\f{1}{4\pi G}\, \oint_S f\, [\Im \Psi_2]\,
\d^2 V\, . \ee
Here $f$ is a `potential' for $\varphi^a$ on $\IH$ \,\,
---given by $\varphi^a = \epsilon^{ab} D_b f$--- \,which exists because
$\Lie_\varphi \epsilon_{ab} \= 0$. By the isolated horizon
boundary conditions it follows that if $\varphi^a$ is the
restriction to $S$ of a vector field on $\IH$ satisfying
$\Lie_\l\, \varphi^a \= 0$, then $J^\varphi_S$ is independent of
the 2-sphere cross-section $S$ used in (\ref{J}).

Thus, on any isolated horizon there is a well-defined notion of a
`generalized angular momentum' $J^\varphi_S$, associated with any
divergence free vector field $\varphi^a$ satisfying $\Lie_\l
\varphi^a \= 0$. $\Im\Psi_2$ plays the role of the `{angular
momentum aspect}'. Hence, it is natural to construct the angular
momentum multipoles $J_n$ by rescaling the $L_n$ with appropriate
dimensionful factors. This strategy is supported also by other
considerations. First, since $\Im \Psi_2$ transforms as a
pseudo-scalar under spatial reflections, we will automatically
satisfy the criterion that the angular momentum multipoles should
transform as pseudo tensors. Second, all angular momentum
multipoles would vanish if and only if $\Im \Psi_2 \= 0$ and this
is precisely the condition defining non-rotating isolated horizons
\cite{afk,abl2}. Thus, the
strategy has an overall coherence.

To obtain the precise expression, let us first recall the
situation in magnetostatics in flat space-time. If the current
distribution $j^a$ is axi-symmetric, the $n$th magnetic moment
$\textbf{M}_n$ is given by:
\be  \textbf{M}_n = \int r^n P_n(\cos \theta)\,\, \vec\nabla\cdot
(\vec{x} \times \vec{j})\, d^3x\, , \ee
where $P_n$ are the Legendre polynomials. If the current
distribution is concentrated on the sphere $S$ defined by $r= R$,
the expression simplifies to:
\be  \label{mag}\textbf{M}_n = - R^{n+1}\, \oint_S (\t\epsilon^{ab}
\t{D}_b P_n(\cos \theta))\,\, \t{j}_a\, d^2V\, , \ee
where $\t\epsilon_{ab}$ is the alternating tensor on the $r=R$
2-sphere and $\t{j}_b$ is the projection of $j_b$ on this
2-sphere. Note that this expression refers just to the
axi-symmetric structure on the 2-sphere $S$ and not to the flat
space in which it is embedded. Comparison of (\ref{J}) with
(\ref{mag}) suggests that we can think of the horizon $\IH$ as
being endowed with a surface `current density'
\be (\t{j}_\IH)_a = \f{1}{8\pi G}\, \tw_a  \ee
and define the angular momentum (or `current') moments as:
\ba \label{angular}
 J_n &= & - \f{R_\IH^{n+1}}{8\pi G}\,\oint_S(\t\epsilon^{ab}
\t{D}_b P_n(\zeta))\, \tw_a\, d^2V \nonumber\\
 &= & - \sqrt{\f{4\pi}{2n+1}}\, \f{R_\IH^{n+1}}{4\pi G}\,\,
 \oint_S Y_n^0 (\zeta) \Im\Psi_2\, d^2 V\nonumber\\ &= &
\sqrt{\f{4\pi}{2n+1}}\, \f{R_\IH^{n+1}}{4\pi G}\,\, L_n\\
 \ea

Let us now turn to the mass multipoles, $M_n$. When all $J_n$
vanish, we should be left just with $M_n$. These are then to be
obtained by rescaling the multipole moments $I_n$ by appropriate
dimensionful factors. In electrostatics, when the charge density
is axi-symmetric, the electric multipoles are defined by
\be \textbf{E}_n = \int r^n P_n(\cos \theta)\, \rho\, d^3x\, .
 \ee
When the charge is concentrated on the sphere $S$ defined by $r=
R$, the expression simplifies to:
\be  \label{ele}\textbf{E}_n =  R^{n}\, \oint_S P_n (\cos\theta)
\t\rho\, d^2V \ee
where $\t\rho$ is the surface charge density. Again, the final
expression refers only to the axi-symmetric structure on the
2-sphere $S$ and not to the flat space in which it is embedded.
Hence we can take it over to type II horizons. What we need is a
notion of a `surface mass density'. Now, Hamiltonian methods have
provided a precise definition of mass $M_\IH$ of type II isolated
horizons in the Einstein-Maxwell theory \cite{afk,abl2}. The
structure of geometric multipoles $I_n$ now suggests that we
regard $M_\IH$ as being `spread out' on the horizon, the `surface
density' $\t\rho_\IH$ being uniformly distributed in the spherical
case but unevenly distributed if the horizon is distorted. It is
then
natural to set%
\footnote{Note incidentally that the smaller the principal radii
of curvature of the intrinsic geometry, the higher is
$\t\rho_\IH$. Thus, the situation has a qualitative similarity
with the way charge is distributed on the surface of a conductor.}
\be \t\rho_\IH = \frac{1}{8\pi}\, M_\IH \tR = -\f{1}{2\pi}\, M_\IH
\Re\Psi_2\, .\ee
This heuristic picture motivates the following definitions:
\ba \label{mass}M_n &:=& - \sqrt{\f{4\pi}{2n +1}}\,\f{M_\IH
R_\IH^n}{2\pi} \oint_S
Y_n^0(\zeta)\, \Re\Psi_2\, d^2V \nonumber\\
&=& \sqrt{\f{4\pi}{2n +1}}\, \f{M_\IH R_\IH^n}{2\pi}\, I_n\, .\ea
Here $M_\IH$ is the isolated horizon mass, which is determined by
the horizon radius $R_\IH$ and angular momentum $J_1$ via:
\be \label{mass1} M_\IH = \frac{1}{2 G R_\IH} \sqrt{R_\IH^4 + 4G^2
J_1^2} \, ,\ee

Since our definitions are based on analogies with source
multipoles in the Maxwell theory, an important question is whether
they have the physical properties we expect in general relativity.
We have the following:
\begin{itemize}

\item As discussed in section \ref{s3.1}, the geometrical
multipoles $L_0$ and $I_1$ vanish. Hence it follows that the
angular  momentum monopole moment $J_0$ vanishes as one would
expect on physical grounds, and the mass dipole moment $M_1$
vanishes implying that we are in the center of mass frame.

\item By construction, the mass monopole $M_0$  agrees with the
horizon mass $M_\IH$ and, by inspection, the angular momentum
dipole moment $J_1$ equals the horizon angular momentum $J_\IH$,
calculated through Hamiltonian analysis \cite{abl2}.

\item Restrictions on the multipoles imposed by symmetries of the horizon 
geometry follow immediately from the remark at the end of section
\ref{s3.1}:\\
i) if the horizon geometry is such that $\Psi_2 \mapsto \Psi_2^*$
under the reflection $\zeta \mapsto -\zeta$, then $M_n =0$ if $n$
is odd and $J_n =0$ if $n$ is even. This is in particular the case
for the Kerr isolated horizon. \\
ii) If $\Im\Psi_2 \=0$, then all angular momentum multipoles
vanish. This is in particular the case if $\IH$ is a Killing
horizon in a static space-time. \\
iii) If the horizon geometry is spherically symmetric, $M_n =0$
for all $n>0$ and $J_n =0$ for all $n$.

\item Our multipoles $M_n, J_n$ are constructed just from the
knowledge of the horizon geometry; knowledge of the space-time
metric in the exterior region is not required. In particular,
there may well be matter sources outside the horizon, responsible
for its distortion and the exterior geometry need not even be
stationary or asymptotically flat. Even when the exterior
\emph{is} stationary and asymptotically flat, there is no a priori
reason to expect that these `source multipoles' would agree with
the `field multipoles' defined at infinity ---unless symmetry
principles are involved--- because the gravitational field outside
the horizon would also act as a source, contributing to the
`total' moments at infinity.

It is interesting to analyze the situation in Kerr space-times.
Because of symmetries, the horizon mass monopole and angular
momentum dipole agree exactly with the corresponding field moments
at infinity. What about the mass quadrupole $M_2$ or the angular
momentum octupole? As figure 1 shows, while the field quadrupole
is not exactly equal to the horizon quadrupole, the difference is
insignificant for small values of the Kerr parameter $a= J/M$ .
Furthermore, even in the extreme limit $a =M$, there is only a
factor of 1.4 between the two mass quadrupoles and 1.14 between
the angular momentum octupoles. In the large $a$ limit, the
effects due to dragging of inertial frames is so large that
post-Newtonian and post-Minkowskian approximations fail; in this
regime, there are no reliable calculations relating the source and
field moments. So a priori it would not have been surprising if
there were a factor of a million!
\begin{figure}[h]
\unitlength=1cm
\begin{minipage}[b]{16.0cm}
\begin{picture}(15.0,7.0)
\includegraphics{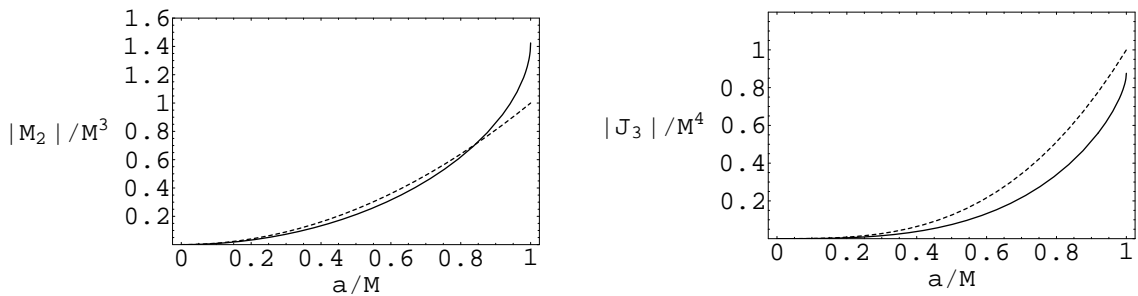}
\end{picture}
\caption{Plots of the horizon multipoles (solid lines) and Hansen's
field multipoles (dashed lines) for Kerr space-times as functions
of $a/M$. The first figure shows the behavior of $|M_2|/M^3$ where
$M_2$ is the mass quadrupole moment and the second shows the
behavior of $|J_3|/M^4$ where $J_3$ is the angular momentum
octupole. Absolute values are used because of the difference in
Hansen's and our sign conventions. (His mass monopole is $-M$
while ours is $M$.) Quantities plotted are dimensionless in the
$G=1$ units.}
\end{minipage}
\label{figure1}
\end{figure}

\item As noted in the Introduction, in vacuum, stationary
space-times Hansen's field moments suffice to
determine the geometry in the neighborhood of infinity. Is
there an analogous result for the horizon multipoles? The answer
turns out to be affirmative. Fix a cross section $S$ of $\IH$ and
consider the future directed, inward pointing null vector field
$n^a$ which is orthogonal to $S$, normalized so that $\ell^an_a =
-1$. The null geodesics originating on $S$ with tangent vector
$-n^a$ generate a null surface $\mathcal{N}$. In the source-free
Einstein theory, there is a well-defined initial value problem
based on the double null surfaces, $\IH$ and ${\cal N}$
\cite{frl}. The freely specifiable data consists of the pair
$(\tq_{ab}, \tw_a)$  on $\IH$ and the Newman-Penrose component
$\Psi_4$ of the Weyl tensor on $\mathcal{N}$ \cite{prl}.

Now suppose that the space-time is analytic near $\IH$ and admits
a stationary Killing field $t^a$ in the neighborhood of $\IH$
which is time-like in the exterior region and becomes null on
$\IH$. Then, as one might imagine, the initial value problem
becomes highly constrained: $\Psi_4$ on $\mathcal{N}$ is
determined by the horizon geometry \cite{abl3}. Hence, the horizon
multipoles suffice to determine the solution to Einstein's
equation in a past neighborhood of $\IH \cup \mathcal{N}$. Thus,
qualitatively the result is the same as with the field multipoles.%
\footnote{However, while the domain in which the solution is
determined by field multipoles is known to be large \cite{bs}, the
available results on the double-null initial value problem only
ensure the existence of the solution in a small, past neighborhood
of $\IH\cup \mathcal{N}$. But it is quite possible that the
double-null initial value results can be significantly
strengthened.}
\end{itemize}

To summarize, the mass and angular momentum multipoles $M_n, J_n$
have physically expected properties. This in turn strengthens the
heuristic picture we used to fix the dimensionful rescalings of
$I_n, L_n$. We first considered stationary, axi-symmetric charge
and current distributions with support on a 2-sphere in the
Maxwell theory and expressed the electric and magnetic multipoles
using only the axi-symmetric structure on the 2-spheres without
reference to Minkowski space-time. We then noted that these
structures are available also on type II horizons.%
\footnote{Note that the spherical harmonics $Y_n^0(\zeta)$ used in
the definition of geometric multipoles $I_n, L_n$ refer to the
unique round metric $\tq^o_{ab}$ defined by coordinates $(\zeta,
\phi) $ (see (\ref{qo})). They are eigenstates of the Laplacian
defined by $\tq^o_{ab}$ and not of the physical metric $\tq_{ab}$.
Therefore, the extension of the Maxwell formulas is natural. As
noted in section \ref{s2.2}, $\tq^o_{ab}$ is \emph{uniquely}
determined by the axi-symmetry of $\tq_{ab}$ and the two metrics
have the \emph{same area element}. Therefore,  $\t\rho_\IH$ and
$(\t{j}_\IH)_a$ can be interpreted as the mass and current
densities in terms of \emph{either} metric.}
Additional structures made available on these horizons by
geometric and Hamiltonian methods then led us to our definitions
of $M_n, J_n$. The physical picture is that observers in the
exterior region (between the horizon and infinity) can regard the
horizon multipoles as arising from an effective (but fictitious)
mass density $\t\rho_\IH = -(1/2\pi) M_\IH \Re \Psi_2$ and a
current density $(\t{j}_\IH)_a = (1/8\pi G) \tw_a$ on $\IH$. This
picture may be useful in physical applications.

\subsection{Extensions}
\label{s3.4}

So far, we have restricted ourselves to non-extremal, type II
isolated horizons which have no matter fields on them. In this
sub-section, we will extend our results in various directions.

\subsubsection{Maxwell fields}
\label{s3.4.1}

Let us restrict ourselves to the non-extremal case but allow
Maxwell fields on $\IH$. Then the Einstein-Maxwell horizon
geometry consists of the quadruplet $(q_{ab}, \D, B_{ab}, E_{ab})$.
The presence of matter fields on $\IH$ causes a few minor
modifications in the discussion of gravitational multipoles. We
will first discuss these and then turn to the electromagnetic
multipoles.

In the gravitational case, the definition of the basic complex
field $\Phi_\IH$ on $\IH$ continues to be the same in terms of
$\tR$ and $\Im \Psi_2$ but changes if we use $\Re \Psi_2$ in place
of $\tR$:
\be \Phi_\IH  := \f{1}{4} \tR - i \Im \Psi_2 = - \Psi_2 +\f{1}{4}
R_{ab} \tq^{ab} . \ee
where the term involving the space-time Ricci tensor can be
expressed in terms of $B_{ab}$ and $E_{ab}$ as:
\be R_{ab} \t{q}^{ab} = G\, (\tB_{ab} \tB^{ab} + \tE_{ab}
\tE^{ab})\, .\ee
The definition of mass $M_\IH$ changes from (\ref{mass1}) to
\be M_\IH = \frac{\sqrt{(R_\IH^2 + GQ_\IH^2)^2 + 4G^2 J_\IH^2}}
{2GR_\IH}\, .\ee
The subsequent definition of the geometric and physical multipole
moments is the same as in section \ref{s3.1}. The reconstruction
of the free data from multipoles is also unaffected but there is
an additional term involving $B_{ab}$ and $E_{ab}$  in the
reconstruction of $\D$ from the free data (see Appendix A).
Finally, there is a minor modification in the list of properties
of $M_n, J_n$ listed in section \ref{s3.3}.  This arises because,
in presence of a Maxwell field on the horizon, the canonical
angular momentum $J_\IH$ obtained by Hamiltonian methods contains
two terms, a gravitational one and an electromagnetic one
\cite{abl2}. The angular momentum dipole moment $J_1$ yields just
`the gravitational part' of $J_\IH$. There does not seem to be a
natural generalization of the definition of angular momentum
multipoles such that $J_1$ would agree with the full $J_\Delta$.

Let us now turn to the electromagnetic fields. As noted in section
\ref{s2.1}, the electromagnetic free data consists of the
projections $\tB_{ab}$ and $\tE_{ab}$ of $B_{ab}$ and $E_{ab}$ on
a cross-section $\t\IH$. Therefore, it is natural to define the
electromagnetic counterpart of $\Phi_\IH$:
\be \Phi^{\rm EM}_\IH :=  - \f{1}{2}\, \t\epsilon^{ab} \,[\tE_{ab}
+ i\t{B}_{ab}] \ee
and define multipoles via:
\ba Q_n &=& \frac{R^n}{\sqrt{4\pi(2n+1)}} \oint_{\t\IH}
{\Re}(\Phi^{\rm EM}_\IH)\, Y_n^0(\zeta)\, \d^2\t{V}\\
P_n &=& \frac{R^n}{\sqrt{4\pi(2n+1)}} \oint_{\t\IH}
{\Im}(\Phi^{\rm EM}_\IH)\, Y_n^0(\zeta)\, \d^2\t{V}\, .
\\ \ea
Clearly, $Q_0$ and $P_0$ are the electric and magnetic charges of
the horizon. Thus, heuristically $\Re (\Phi^{\rm EM}_\IH)/4\pi$
and ${\Im}(\Phi^{\rm EM}_\IH)/4\pi$ may be thought of as `surface
charge densities' on the horizon and charge multipoles capture the
non-uniformity in the distributions of electric and magnetic
charge densities.

\textsl{Remark:} If non-electromagnetic sources are also present,
we can still define the gravitational (and electromagnetic)
multipoles as above but the multipoles for other sources have to
be defined case by case. Gravitational multipoles again determine
the `free data' for the horizon geometry $(q_{ab}, \D)$. However,
to reconstruct the horizon geometry from this data, one needs to
know those matter fields which determine the components 
$\tq_a{}^c\, \tq_b{}^d \, R_{cd}$ of the Ricci
tensor, evaluated at the horizon.

\subsubsection{Extremal horizons}
\label{s3.4.2}

Let us now summarize the situation with extremal horizons. In this
case, important modifications are needed in the gravitational
sector. As noted in section \ref{s2.1}, surprisingly, $(q_{ab},
\omega_a, \tB_{ab}, \tE_{ab})$ are now \emph{universal},
determined by the Kerr-Newman family \cite{lp}. However, they no
longer suffice to determine the derivative operator $\D$ on the
horizon because the second rank, symmetric tensor field $\tS_{ab}
= \tq_a{}^c\tq_b{}^d\, \D_c n_d$ is now free \cite{abl1} (see
Appendix A). Therefore, multipole moments characterizing the free
data in the geometry have to be constructed from components of
$\tS_{ab}$. Again, as with $\tw_a$, there is a gauge freedom in
$\tS_{ab}$ associated with the choice of the cross-section $\t\IH$
used to evaluate it. This can be eliminated by fixing the trace of
$\t{S}_{ab}$ \cite{abl3}. Then the gauge invariant information is
coded in the two components of the trace-free part of
$\t{S}_{ab}$.

Of course, we can still use $\Phi_\IH$ to introduce the geometric
multipoles $I_n,L_n$ and electromagnetic multipoles $Q_n, P_n$.
But they are universal; the same as in the extremal Kerr-Newman
case. However, now there are \emph{two additional sets of
geometric multipoles} obtained by integrating the two independent
components of $\tilde{S}_{ab}$ against $Y^0_n(\zeta)$. They
capture the free data which can vary from one extremal isolated
horizon to another. To our knowledge, they do not have a simple,
direct physical interpretation.

\subsubsection{Type III horizons}
 \label{s3.4.3}

The notion of the horizon geometry and the calculations leading to
the free data do not refer to the axial Killing field at all
\cite{abl1} (see Appendix A). In the above, we restricted
ourselves to type II horizons primarily because they are the ones
which are physically most interesting. On a non-extremal type III
horizon, the complex-valued seed function $\Phi_\IH$ again
captures the gauge invariant part of the free data.%
\footnote{In the extremal case, further work is needed to
understand the constraint on $(\tq_{ab}, \t\omega_a)$ which, in the
type II case led to the surprising result that these fields must
coincide with those on an extremal Kerr isolated horizon.}
However, in absence of axi-symmetry, we no longer have the natural
coordinate $\zeta$.\, $\Phi_\IH$ is now a function of two
coordinates on $\t\IH$. Nonetheless, we can use the Laplacian
operator $\tq^{ab} \t\D_a\t\D_b$ on $\t\IH$ to decompose
$\Phi_\IH$ in to generalized spherical harmonics. Thus, now the
geometric multipoles $I_{n,m}, L_{n,m}$ are labelled by two
integers where $n$ labels the discrete eigenvalue of the Laplacian
and $m$ is the degeneracy label of eigenvectors with same
eigenvalue $n$ of the Laplacian. Should a physically interesting
application of type III horizons arise, it should not be difficult
to extend this basic framework further.

There is, however, an important subtlety. In the type II case, the
presence of axi-symmetry led us to single out a preferred function
$\zeta$ and our multipoles were defined using the $Y_n^0(\zeta)$
associated with this $\zeta$. As noted in section \ref{s3}, these
$Y_n^0$ are spherical harmonics associated with the round 2-sphere
metric $\tq^o_{ab}$ of (\ref{qo}). In general, they are \emph{not}
eigenfunctions of the Laplacian of the physical metric $\tq_{ab}$!
Our procedure was essential to ensure that the angular momentum
dipole $J_1$ agrees with the horizon angular momentum $J_\IH$
defined by Hamiltonian methods \cite{abl2}. Had we used multipoles
through spherical harmonics associated with $\tq_{ab}$, this
equality would not have held in general. Thus the extra structure
available on type II horizons played an important role in our
definition of multipoles in sections \ref{s3.1} and \ref{s3.3}.
Since it is absent on a general type III horizon, multipoles
defined using the physical Laplacian will not have all the
physically desirable properties discussed in section \ref{s3.3}.

\section{Discussion}
\label{s4}

Source multipoles in Newtonian gravity characterize the way in
which mass is distributed, higher multipoles providing a complete
description of distortions, i.e., departures from sphericity.
Multipole moments defined in this paper do the same for the
horizon geometry. In this sense, they can be regarded as `source
multipoles' associated with a black hole, distinct from the `field
multipoles' defined at infinity. In Newtonian gravity and
Maxwellian electrodynamics, there is a simple relation between the
two because both these theories are `Abelian': the field does not
serve as its own source. In non-Abelian contexts such as
Yang-Mills theory and general relativity, the field in the region
between the source and infinity itself acts as an effective
source. Hence one would expect there to exist two \emph{distinct}
sets of multipoles, one associated with the central gravitating body
and the other associated with the entire system. If space-time is
stationary, one would expect a simple relation between mass
monopoles, if it is axi-symmetric, between angular momentum
dipoles and if in addition there is reflection symmetry, one would
expect all odd mass multipoles and even angular momentum
multipoles to vanish in \emph{both regimes}. However, in absence
of symmetries, one does not expect a simple relation between the
two sets in the fully relativistic regimes where post-Newtonian
and post-Minkowskian approximations fail. This point seems to have
been glossed over in the general relativity literature on
applications of multipoles. Since horizon multipoles were
unavailable, one typically identified the field multipoles with
the multipoles of the black hole and used them in \emph{all}
contexts.

The horizon multipoles defined in this paper are likely to have
three sets of applications which we now summarize.

The first is to the problem of equations of motion. In the more
recent literature on this subject, multipoles have been used in two different
contexts. The first is illustrated by the problem of extreme
mass-ratio binaries where a solar mass black hole orbits about a
central supermassive one. For this case, there is an interesting
framework \cite{fr} enabling one to use gravitational wave
observations to `measure' the gravitational multipole moments of
the central object. In actual calculations, one typically assumes
that the small hole is far from the supermassive one. One can then
express the space-time metric at the position of the small hole as
a series in inverse powers of distance from the central hole using
the Hansen field-multipoles  as coefficients, and calculate
quantities of physical interest. Within the approximations
inherent to these calculations, it is entirely appropriate to use
the field multipoles and the framework provides a way of measuring
\emph{these} multipoles from the gravitational wave data. However,
the approximation will fail if the orbit lies in a genuinely
strong field regime. In that case, one may be able to use the near
horizon expansion of the metric \cite{prl}. Here expansion
coefficients refer to the horizon geometry but it should be
straightforward to recast them in terms of the horizon multipoles.

The second context is illustrated in the study of how the
gravitational field of \emph{other} objects affects a given, small
black hole (see, e.g., \cite{tpm}). Calculations of the resulting
distortion, precession, spin-flips, etc. of the black hole require
the knowledge of tidal forces and torques exerted by other bodies
on the black hole. Typically, these are calculated from
interaction terms involving a coupling of black hole multipoles to
perturbations in the Weyl curvature caused by other objects (at
the location of the black hole). In these calculations it is
inappropriate to use the \emph{field} multipoles of the small
black hole since the coupling occurs at the location of the black
hole. What one needs is multipoles associated with the horizon,
not coefficients in the asymptotic expansion of the metric
produced by the small black hole alone. The horizon multipoles
introduced here provide the appropriate notion.

A second set of applications is to binary black hole simulations
in numerical relativity. At very early times when the two black
holes are widely separated and at late times when the single black
hole has settled down, the world tubes of apparent horizons are
well approximated, within numerical errors, by isolated horizons.
One often wants to compare results of two different numerical
simulations, particularly at late times. These comparisons are
generally difficult in the strong field regime because the results
are tied to the coordinates used. Our mass and angular momentum
multipoles provide a diffeomorphism invariant method to compare
such results. Furthermore, when there are differences, they
provide a tool to interpret them physically.

A more interesting potential application is in the dynamical
regimes, where the world tubes of apparent horizons can be
modelled by dynamical horizons $H$ \cite{ak}. These are naturally
foliated by marginally trapped 2-spheres $S_H$. On each $S_H$,
there is an obvious, well-defined analog of $\Phi_\IH$. Therefore,
a natural strategy is to use it to define multipoles which would
now be \emph{time-dependent}. It would be very interesting to
analyze the evolution of these moments both analytically and
numerically. For example, there is a well-defined notion of energy
flux across any $S_H$ \cite{ak}. Is there then an analog of the
quadrupole formula \emph{at the horizon}? From the knowledge of
the horizon quadrupole and its relation to the Kerr quadrupole,
can one gain insight into the maximum amount of energy that can be
emitted in gravitational radiation? For non-rotating black holes,
early numerical simulations \cite{earlynr} introduced a notion of
horizon multipoles, using tools then available. They then
calculated a complex frequency describing the time evolution of
the quadrupole, where the imaginary part of the frequency captures
damping. They found that this frequency coincided with the
quasi-normal frequency of the final black hole. Although the
calculation of multipoles was somewhat imprecise, the result is simple
and intriguing. Interestingly, our horizon multipoles provide a
precise formulation of the physical ideas underlying those more
intuitive notions of multipoles. One is therefore led to ask: Can
one analyze the final stages of mergers using the present,
diffeomorphism invariant multipoles and show in detail that the
complex frequency is the same as the quasi-normal one? Such a
result would provide a deep insight in to the relation between the
highly non-linear dynamics of the horizon and the quasi-normal
ringing associated with the linear regime. Again, the issue can be
examined using analytical methods as well as numerical
simulations.

The final set of applications is to quantum gravity. To calculate
black hole entropy from first principles, one needs to construct
an ensemble, where the `macroscopic parameters' describing the
system are fixed. To be physically meaningful, these parameters
have to be diffeomorphism invariant. For type I horizons, this is
straightforward: there is only one gravitational parameter, which
can be taken to be the horizon area (or mass). For type II
horizons, one can not just work with mass and angular momentum
because the horizon may be distorted by various types of hair.
Even within the 4-dimensional Einstein-Maxwell theory where
no-hair theorems hold, distortions can be caused by matter rings
and other black holes. Even if the black hole itself is isolated,
one can not automatically rely on uniqueness theorems and say that
it must be a Kerr hole because it is physically unreasonable to
require that the whole space-time is stationary. Since one wishes
to calculate entropy of a black hole in equilibrium, one should
only need to require that the horizon geometry is time
independent, not the whole universe. To incorporate physically
realistic situations, then, one needs a diffeomorphism invariant
characterization of the horizon fields. Multipoles can now serve
as the required parameters in the construction of the ensemble. It
turns out that they can in fact be used to calculate entropy
associated with type II horizons \cite{entropy}.

\section*{Acknowledgments}
We would like to thank Piotr Chrusciel and John Stachel for
useful correspondence, and Eanna Flanagan, Jerzy Lewandowski and Ben Owen 
for helpful
discussions. This work was supported in part by the National
Science Foundation grants PHY-0090091, PHY99-07949, the National
Science Foundation Cooperative Agreement PHY-0114375, the
Eberly research funds of Penn State and the Polish Committee
for Scientific Research (KBN) grant 2 PO3B 130 24.

\begin{appendix}

\section{Reconstruction of the horizon geometry from the free
data}

For convenience of the reader, in this Appendix we will collect
some results from \cite{abl1} which play an important role in this
paper. Specifically, we will show how the horizon geometry
$(q_{ab}, \D)$ is determined by the free data via field equations.

Fix a 3-manifold $\IH$, topologically $\S^2\times \R$ and a
2-sphere cross-section $\t\IH$ thereon. We wish to show that,
given a pair of fields $(\tq_{ab}, \tw_a)$ on $\t\IH$, there is a
non-extremal isolated horizon structure $([\l^a], q_{ab}, \D)$, unique
up to diffeomorphism, on
$\IH$ such that $\tq_{ab}$ is the projection of $q_{ab}$ on
$\t\IH$;\, $D_a \l^b = \omega_a \l^b$;\, $\tw_a$ is the projection
of $\omega_a$ on $\t\IH$ and $\omega_a \ell^a \not=0$. (We will
turn to the extremal case at the end.)

Introduce a function $u$ whose level surfaces provide a foliation
of $\IH$ by a family of 2-spheres such that $\t\IH$ is a leaf. Let
$n_a = -D_au$ and define a vector field $\l^a$ such that $\l^an_a
\=-1$. Thus, $\l^a$ is transversal to the foliation. Now, given
$\tq_{ab}$ on $\t\IH$, there is a unique tensor field $q_{ab}$ on
$\IH$ such that :
 \be q_{ab} \l^b =0,\quad \Lie_\l q_{ab} =0, \quad {\rm and} \quad
q_{ab} \t{V}^a \t{W}^b = \tq_{ab} \t{V}^a \t{W}^b \ee
for all vector fields $\t{V}^a, \t{W}^a$ tangential to $\t\IH$.
This is the required $q_{ab}$.

The $\D$ we seek is to be the derivative operator on a
non-extremal isolated horizon. So, it must satisfy: i) $\D_a
q_{bc} \=0$, ii) $[\Lie_\l,\, \D] \=0$, and iii) $D_a\l^b = \omega_a
\l^b$ for some 1-form $\omega_a$ where $\l^a \omega_a = \k_\l$ is
non-zero. Consider first torsion-free derivative operators $\D$ on
$\IH$ which are compatible with $q_{ab}$. They all have the same
action on 1-forms $f_a$ satisfying $f_a \l^a =0$, given by:
\be \D_a f_b = \f{1}{2}\, \Lie_{\bar f}\, q_{ab} + \D_{[a}
f_{b]}\ee
where ${\bar f}^a$ is any vector field on $\IH$ satisfying ${\bar
f}^a q_{ab} = f_b$. (While ${\bar f}^a$ is not unique, one can
check that the first term on the right side is insensitive to this
ambiguity.) What distinguishes these derivative operators is only
their action on $n_a$. Set
\be S_{ab} :=\D_a n_b \ee
Then, $S_{ab}$ is symmetric and $\l^b S_{ab} = \l^b \D_a n_b =
\omega_a$. Thus, the freedom in the definition of $\D$ is
completely captured by the pair $(\omega_a, \tS_{ab})$ where
$\tS_{ab}$ is the projection of $S_{ab}$ on the leaves of the
foliation (defined by the projection operator $\tq_a{}^b
=\delta_a{}^b + n_a\l^b$).

We now bring in the pull-backs of the field equations to $\IH$. A
simple calculation yields:
 \be  \underbar{R}_{ab} \l^b \= 2 \l^a \D_{[a} \omega_{b]}\, ,\ee
where the underline denotes the pull-back to $\IH$. Now,
conditions on the stress-energy tensor in the definition of a
non-expanding horizon imply that $\underbar{T}_{ab}\l^a\= 0$
whence $\underbar{R}_{ab}\l^b \= 0$. Therefore, via the Cartan
identity, this part of the Einstein equation is equivalent to:
\be \D_a (\omega_b\l^b) - \Lie_\l \omega_a \=0 \ee
Now, since $[\Lie_\l, \, \D] \=0$, it follows that $\Lie_\l
\omega_a \=0$, whence $\k_\l$ is constant on $\IH$. The remaining
projection of Einstein's equations on $\IH$ determines how
$S_{ab}$ `evolves' along $\IH$:
\be \label{ls} \Lie_\l \, \tilde{S}_{ab} \= -\k_l \tilde{S}_{ab} +
\tilde\D_{(a} \tilde\omega_{b)} + \tilde\omega_a \tilde\omega_b -
\frac{1}{2} \tilde{\cal R}_{ab} + \frac{1}{2} \tq_a{}^c\,
\tq_b{}^d \, R_{cd} \ee
where $\tilde{\D}$ and $\tilde{\cal R}_{ab}$ denote the derivative
operator and the Ricci tensor on the leaves of the foliation and
$\tilde \omega_a$ is the projection of $\omega_a$ on these
cross-sections. However, since $[\Lie_\l, \, \D] \=0$, the left
side vanishes. Hence, \emph{if} $\k_\l \not= 0$, we have:
\be \tS_{ab} = \f{1}{\k_\l}\left[\,\tilde\D_{(a} \tilde\omega_{b)}
+ \tilde\omega_a \tilde\omega_b - \frac{1}{2} \tilde{\cal R}_{ab}
+ \frac{1}{2} \tq_a{}^c\, \tq_b{}^d \, R_{cd}\right]\, . \ee
Thus, given the part of the stress-tensor which determines the
projection $\tq_a{}^c\, \tq_b{}^d \, R_{cd}$ of the space-time
Ricci tensor on the leaves of the foliation, $\tS_{ab}$ is
completely determined by $\tq_{ab}$ and $\tw_a$. Since $\Lie_\l
q_{ab} \=0$ and  $[\Lie_\l, \D] \=0$, it follows that it suffices
to specify $(\tq_{ab}, \tw_a)$ only on one leaf of the foliation.

To summarize, in the non-extremal case, the field equations
pulled-back to $\IH$ imply that $\D$ is completely determined by
$\tq_{ab}, \tw_a, \k_\l$ and $\k_\l$ is a constant on $\IH$. An
examination of the other field equations shows that they do not
constrain the horizon geometry; rather, they provide evolution
equations off $\IH$. Thus, given $(\tq_{ab}, \tw_a)$ on $\t\IH$,
we can construct a triplet $([\l], q_{ab}, \D)$ on $\IH =
\t\IH\times \R$ providing a non-extremal isolated horizon
geometry. All such possible triplets are diffeomorphically related
because the only freedom in the construction is the choice
of $[\l]$. (As noted in the main text, the value of $\k_\l$ is not
fixed on an isolated horizon because of the constant rescaling
freedom in $[\l]$.)

Let us now turn to the extremal case. Analysis is identical until
(\ref{ls}). But, since $\k_l$ vanishes, $\tS_{ab}$ decouples from
the equation and remains completely unconstrained. However, we now
have a \emph{constraint} on the pair $(\tq_{ab}, \tw_a)$:
\be \tilde\D_{(a} \tilde\omega_{b)} + \tilde\omega_a
\tilde\omega_b - \frac{1}{2} \tilde{\cal R}_{ab} + \frac{1}{2}
\tq_a{}^c\, \tq_b{}^d \, R_{cd} = 0\ee
This equation is quite complicated and its solutions are not known
in the general case. However, on type II horizons in the
Einstein-Maxwell theory, a surprising simplification occurs: in
this case every solution $(\tq_{ab}, \tw_a)$ to this equation is
diffeomorphically related to that on the Kerr-Newman extremal
horizon \cite{lp}. Hence, the freely specifiable data is just
$\t{S}_{ab}$.

\end{appendix}


\begin{thebibliography}{99}

\bibitem{ae} Einstein A 1916 N\"aherungsweise Integration der
Feldgleichungen der Gravitation \textit{Sitzungsberichte der
K\"oniglich Preussischen Akademie der Wissenschaften} (Berlin)
688-696.

\bibitem{sb} Sachs R K and Bergmann P G 1958 Structure of particles
in linearized gravitational theory \textit{Phys.
Rev.} \textbf{112}, 674-680

\bibitem{gho}
Geroch R 1970 Multipole moments. II. curved space \textit{J. Math. Phys.}
\textbf{11} 2580-2588 \\
Hansen R 1974 Multipole moments in stationary space-times
\textit{J. Math. Phys.} \textbf{15} 46-52\\
Beig R and Simon W 1983 The multipole structure of stationary
space-times \textit{J. Math. Phys.} \textbf{24}, 1163-1171

\bibitem{bs}
Beig R and Simon W 1980 Proof of a multipole conjecture due to Geroch
\textit{Commun. Math. Phys} \textbf{78}
75-82 \\
Beig R 1981 The multipole expansion in general relativity
\textit{Acta Phys. Aust.} \textbf{53}
249-270\\
Beig R and Simon W 1981 On the multipole expansion of stationary
space-times \textit{Proc R. Soc.} (London) \textbf{A376} 333-341

\bibitem{gd} Dixon G 1979 Extended bodies in general relativity:
their description and motion, in \textit{Isolated gravitating
systems in general relativity} ed Ehlers J (Amsterdam: North
Holland)

\bibitem{recent} Anandan J, Dadhich N and Singh P 2003 Action 
principle formulation for the motion of extended bodies 
in general relativity \emph{Phys.
Rev.} \textbf{D68} 124014\\
Ohashi A 2003 Multipole particle in relativity \emph{Phys. Rev.}
\textbf{D68} 044009

\bibitem{prl} Ashtekar A,  Beetle C, Dreyer O, Fairhurst S,
Krishnan B, Lewandowski J and Wi\'sniewski J 2000 Generic isolated
horizons and their applications \textit{Phys.\ Rev.\ Lett.}
\textbf{85}, 3564-3567 

\bibitem{abf} Ashtekar  A, Beetle C and Fairhurst S 1999
Isolated horizons: a generalization of black hole mechanics
\textit{Class.\ Quantum Grav.} \textbf{16} L1--L7\\
Ashtekar  A, Beetle C and Fairhurst S 2000 Mechanics of isolated
horizons \textit{Class.\ Quantum Grav.} \textbf{17} 253--298

\bibitem{afk} Ashtekar  A,  Fairhurst  S and Krishnan  B 2000 Isolated
horizons: Hamiltonian evolution and the first law  \textit{Phys. \
Rev.} \textbf{D62} 104025.

\bibitem{abl1} Ashtekar  A, Beetle  C and Lewandowski J 2002
Geometry of generic isolated horizons  \textit{Class. Quantum
Grav.} \textbf{19}  1195--1225

\bibitem{abl2} Ashtekar  A, Beetle  C and Lewandowski  J
2001 Mechanics of rotating isolated horizons \textit{Phys.\ Rev.\ }
\textbf{D64}  044016

\bibitem{frl}Friedrich H 1981 On the regular and asymptotic
characteristic initial value problem for Einstein's field
equations \textit{Proc.\ R.\ Soc.\ } \textbf{375} 169-184\\
Rendall A 1990 Reduction of the characteristic initial value
problem to the Cauchy problem and its applications to Einstein's
equations \textit{Proc.\ R.\ Soc.\ } \textbf{427} 221-239\\
Lewandowski J 2000 Spacetimes admitting isolated horizons
\textit{Class.\ Quantum Grav.\ } \textbf{17} L53-L59

\bibitem{abl3} Ashtekar A, Beetle C and Lewandowski J (2004)
Space-time geometry near isolated horizons (in preparation)\\
Pawlowski, T (2004) Ph.D. Dissertation, University of Warsaw (in
preparation)

\bibitem{acs} Ashtekar  A, Corichi  A and Sudarski D 2003
Non-minimally coupled scalar fields and isolated horizons
\textit{Class. Quantum Grav.} \textbf{20} 3413-3425

\bibitem{pc} Chrusciel P 1992 On the global structure of
Robinson--Trautman space-times
\textit{Proc.\ Roy.\ Soc.\ } \textbf{436} 299-316

\bibitem{lp} Lewandowski J and Pawlowski T 2003 Extremal isolated
horizons: a local uniqueness theorem \textit{Class. Quantum Grav.}
\textbf{20} 587-606

\bibitem{jn} Janis A I and Newman E T 1964 Structure of
gravitational sources \textit{J. Math. Phys.} \textbf{6} 902-914

\bibitem{Date} Date G 2000 Notes on isolated horizons
\textit{Class. Quantum Grav.} \textbf{17} 5025-5046

\bibitem{fr} Ryan F 1995 Gravitational waves from the inspiral of
a compact object in to a massive, axi-symmetric body with
arbitrary multipole moments \textit{Phys. Rev.} \textbf{D52}
5707-5718

\bibitem{tpm} Thorne K S, Price R H and Macdonald D A 1986
\textit{Black Holes: The Membrane Paradigm} (New Haven and London,
Yale UP), chapter 5

\bibitem{ak} Ashtekar A,  Krishnan B 2002 Dynamical horizons:
energy, angular momentum, fluxes and balance
laws \textit{Phys. Rev. Lett.} \textbf{89} 261101\\
Ashtekar A,  Krishnan B 2003 Dynamical horizons and their
properties \textit{Phys. Rev.} \textbf{D68} 104030

\bibitem{earlynr} Anninos P, Bernstein D, Brandt S, Hobill D,
Seidel E, Smarr L 1994  Dynamics of black hole apparent horizons
{\it Phys. Rev. D} {\bf 50} 3801-3819 \\
Anninos P, Bernstein D, Brandt S, Hobill D, Seidel E, Smarr L 1995
Oscillating apparent horizons in numerically generated spacetimes
{\it Aust. J. Phys.} {\bf 48} 1027-1043

\bibitem{entropy} Ashtekar  A 2003 Black hole entropy: inclusion of
distortion and angular momentum
http://www.phys.psu.edu/events/index.html?event\_id=517\&event\_type=17\\
Ashtekar A, Engle J and Van Den Broeck C 2003 (in preparation)

\end{thebibliography}
\end{document}